\documentclass[lettersize,journal]{IEEEtran}
\usepackage{amsmath,amssymb,amsfonts}
\usepackage{algorithmic}
\usepackage{algorithm}
\usepackage{array}

\usepackage[caption=false,font=footnotesize,labelfont=rm,textfont=rm]{subfig}
\usepackage{textcomp}
\usepackage{booktabs}
\usepackage{stfloats}
\usepackage{multirow}
\usepackage{url}
\usepackage{verbatim}
\usepackage{graphicx}
\usepackage{xcolor}
\def\BibTeX{{\rm B\kern-.05em{\sc i\kern-.025em b}\kern-.08em
    T\kern-.1667em\lower.7ex\hbox{E}\kern-.125emX}}
\begin{document}

\title{Channel Gain Map Construction based on Subregional Learning and Prediction\\}

\author{Jiayi Chen,
Ruifeng Gao,
Jue Wang,
Shu Sun,
Yi Wu\thanks{

Jiayi Chen and Jue Wang are with the School of Information Science and Technology, Nantong University, Nantong 226019, China (e-mail: 2230320035@stmail.ntu.edu.cn; wangjue@ntu.edu.cn).

Ruifeng Gao is with the School of Transportation and Civil Engineering, Nantong University, Nantong 226019, China (e-mail: grf@ntu.edu.cn).

Shu Sun is with the Department of Electronic Engineering and the Cooperative Medianet Innovation Center, Shanghai Jiao Tong University, Shanghai 200240, China (e-mail: shusun@sjtu.edu.cn).

Yi Wu is with the College of Photonic and Electronic Engineering, Fujian Normal University, Fujian 350007, China (e-mail: wuyi@fjnu.edu.cn).
}}

\maketitle
\begin{abstract}

The construction of channel gain map (CGM) is essential for realizing environment-aware wireless communications expected in 6G, for which a fundamental problem is how to predict the channel gains at unknown locations effectively by a finite number of measurements. As using a single prediction model is not effective in complex propagation environments, we propose a subregional learning-based CGM construction scheme, with which the entire map is divided into subregions via data-driven clustering, then individual models are constructed and trained for every subregion. In this way, specific propagation feature in each subregion can be better extracted with finite training data. Moreover, we propose to further improve prediction accuracy by uneven subregion sampling, as well as training data reuse around the subregion boundaries. Simulation results validate the effectiveness of the proposed scheme in CGM construction.

\end{abstract}


\section{Introduction}

To support the largely increased data demands and connection requirements, communication network is becoming more complex in the forthcoming 6G era \cite{10054381}. This brings challenges to low-complexity network deployment optimization and transmission design \cite{10616106}. Environment-aware communication provides a promising solution for this challenge, which requires communication-related environment information, also known as the channel knowledge map (CKM) \cite{9373011, drones8050191}, as side information to be exploited when designing the system. In general, the CKM can be presented in terms of a site-specific database, which provides information of concerned channel parameters at given geometric locations. Depending on the particular channel information it conveys, different types of CKM have been studied, including the channel shadowing map \cite{7997333}, channel gain map (CGM) \cite{9771802}, and beam index map \cite{8240727}, etc. Different CKMs can be exploited for different tasks.

Despite the different forms, all CKMs basically form a mapping between possible user locations and channel parameters. To build such a map, electromagnetic simulation techniques like ray tracing can be used. However, ray tracing is computational intensive, especially when the required number of location samples is large. Besides, environment modeling inevitably deviates from the realistic propagation environment, which may degrade its precision. Alternatively, practical measurement can be conducted to obtain more precise and realistic channel data \cite{10210347}. Yet, one cannot expect to construct an entire CKM via measurement as it would be over-costly.

Therefore, a fundamental task in CKM construction is to predict the channels at unknown locations, on the basis of a limited number of known data samples effectively. Conventional spatial interpolation techniques can be used, such as the Kriging \cite{5711699} and inverse distance weighted (IDW) \cite{8610487}. However, the precision cannot be guaranteed in a complex propagation environment and with a size-limited dataset. In addition, some papers constructed the CKM based on model-driven approaches \cite{9771802, 8662745}. Since channel prediction is formulated as a function of distance, it may encounter challenges in adapting to complex environments. Alternatively, a promising approach is to use deep learning methods, which can effectively improve the accuracy of map construction \cite{9354041}. However, these approaches try to characterize the entire propagation environment with a single model. When the concerned environment is complex, using a single prediction model may require a large amount of training data and a complex model structure to achieve satisfactory performance. 


\begin{figure*}[t]
\begin{center}
    \includegraphics[width=1.9\columnwidth]{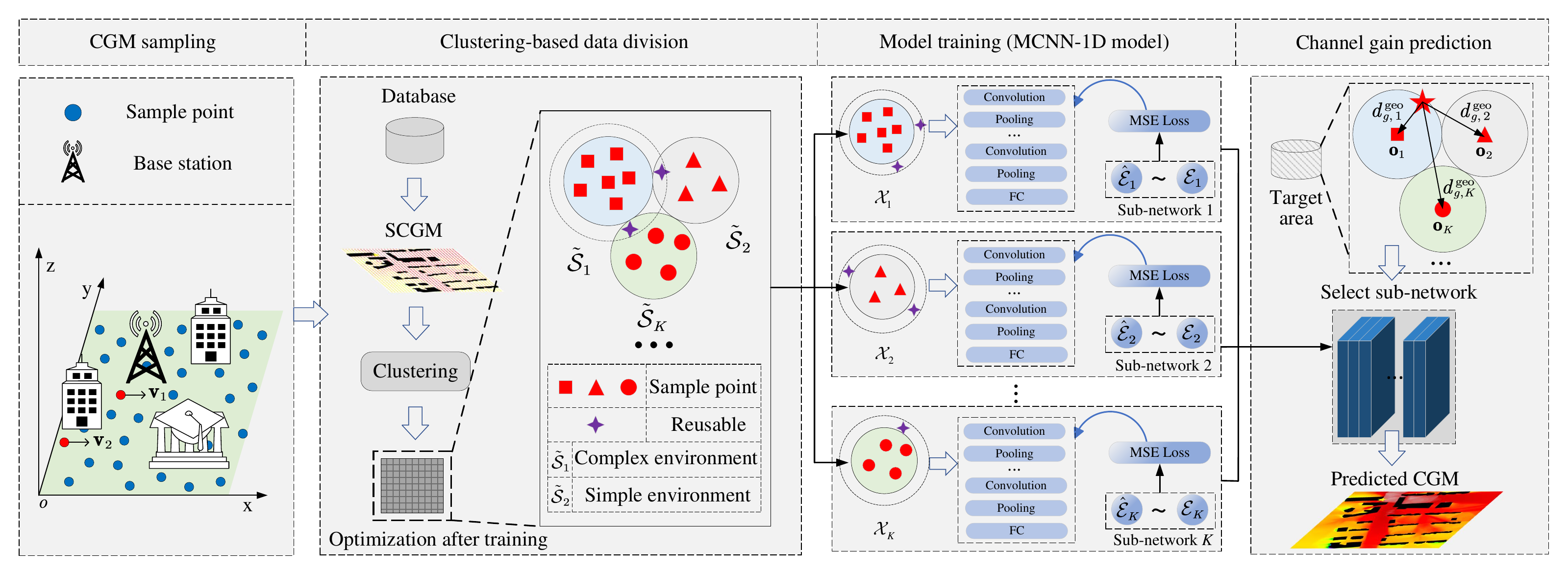}
    \caption{The overall architecture of the subregional learning-based CGM construction.}
    \label{fig1}
\vspace{-4mm}
\end{center}
\end{figure*}

Intuitively, dividing a large map into smaller subregions can better exploit unique propagation properties in different regions. This idea has been adopted in \cite{9771802}, where ground truth sample points were initially grouped based on the expectation maximization (EM) algorithm. Then, the model-based method was employed to predict the channel gain for the target groups. In addition, the work in \cite{8032491} divided the map into line-of-sight (LoS) and non-LoS (NLoS) regions based on ray tracing. However, this map division method requires highly accurate environment modeling and is computationally time-consuming. In complex propagation environments characterized by irregularly distributed obstacles and buildings, applying such a method becomes more challenging.

In this paper, we focus on the construction of CGM, which has been applied in various fields. Traditionally, the CGM was considered static and can either be utilized by the base station (BS) for resource allocation or by mobile users to select the communication modes \cite{10683332}. Zhang \textit{et al.} proposed a CGM-based path planning framework for the unmanned aerial vehicle, utilizing the map to provide large-scale channel gains at uniformly sampled locations on a 3D grid \cite{9269485}. Additionally, the radio map was used for wireless indoor localization, improving accuracy by providing an up-to-date map \cite{8003484}. For the wide usage of CGM, developing an accurate and efficient construction method is critical. The work \cite{10683332} explored both model-based and model-free approaches to construct the CGM. This paper proposes a novel map construction approach based on subregional learning and prediction. The main contributions of this paper are as follows:

\begin{itemize}
\item A data-driven clustering approach is employed to partition the known sample points into clusters and correspondingly form subregions. We perform individual deep learning model construction and training for each subregion to explore their unique environmental characteristics.

\item To further improve the prediction accuracy, we develop an uneven subregion sampling strategy, which calculates the sampling rate for each subregion. Furthermore, we propose a training data reuse scheme to exploit the channel similarity appearing around subregion boundaries.
\end{itemize}

\section{System Model and Problem Description}

We consider the scenario illustrated in the CGM sampling stage of Fig.~\ref{fig1}, where a fixed-position BS is deployed to provide communication service to a target area, for which the CGM is to be constructed to assist system design. For this purpose, the entire area is sampled and each of these samples provides channel-related information at its location.\footnote{In practice, this can be realized by selecting partial of the users to measure the required information and feed it back to the BS.} The collected information is stored in a database, and correspondingly forms a sampled CGM (SCGM). Note that the SCGM is in general geographically sparse, as practical measurement and feedback are costly. Therefore, a prediction model is required to predict channel gains at unknown locations and correspondingly construct the complete CGM for continuous locations in the target area. 

For the $i$-th sample point in the SCGM, the complete information is expressed as $\mathbf{a}_{i}=\left [ \mathbf{c}_{\mathrm{bs} }^T,\mathbf{c}_{\mathrm{sp},i}^T,p,w_i,r_i\right ]^T\in \mathbb{R}^{9\times 1}$, where $\mathbf{c}_{\mathrm{bs} }\in\mathbb{R}^{3\times 1}$ and $\mathbf{c}_{\mathrm{sp},i}\in\mathbb{R}^{3\times 1}$ are the spatial coordinates of the BS and the sample point, respectively, $p\in\mathbb{R}^+$, $w_i\in\mathbb{R}^+$ and $r_i\in\mathbb{R}$ denote the transmit power of the BS, carrier frequency, and average received power, respectively. Note that we primarily focus on large-scale fading in this paper. The $r_i$ at each sample point is derived from simulated data through ray-tracing. Therefore, $r_i$ is fixed. In practice, the received power is typically measured multiple times and averaged to remove the impact of small-scale fading. Denoting the channel gain as $e_i=\frac{r_i}{p}$, and considering that $p$, $\omega_i$, and the sampling height are all fixed for brevity,\footnote{When these parameters vary, the idea can be similarly applied while the correlation between different parameter dimensions can be further exploited.} useful information for CGM construction is extracted from $\mathbf{a}_i$ as 
\begin{equation}
\label{eq1}
\mathbf{v}_{i}=\left [c_{x,\mathrm{bs}},c_{y,\mathrm{bs}},c_{x,i},c_{y,i},e_i\right ]^T\in \mathbb{R}^{5\times 1}, 
\end{equation}
where $c_{x,\mathrm{bs}}$, $c_{y,\mathrm{bs}}$ and $c_{x,i}$, $c_{y,i}$ denote the coordinates of the BS and sample $i$ along the $x$ and $y$ axis, respectively. Preserving the location of the BS facilitates the 
prediction model to extract the relative relationships between sample points and the BS. Then, the SCGM is described as 
\begin{equation}
\label{eq2}
 \tilde {\mathcal{S}}=\{\mathbf{v}_{i}|i=1,..., M_{\mathrm{scgm}}\},   
\end{equation}
with $|\tilde {\mathcal{S}}|=M_{\mathrm{scgm}}$. Now, the task is to construct a continuous CGM estimation, denoted as $\hat{\mathcal{S}}$, given $\tilde{\mathcal{S}}$ as a known dataset.

Using a single model to construct $\hat{\mathcal{S}}$ may be inefficient. In complex scenarios, sampling points at different locations encounter diverse propagation environments \cite{2024fieldfar}. This is illustrated by the example points $\mathbf{v}_1$ and $\mathbf{v}_2$ in Fig.~\ref{fig1}. Clearly, these two points experience different propagation environments, and hence different models should be trained for {\em local CGM prediction}. For this purpose, we partition the SCGM into multiple clusters, each corresponding to a non-overlapping subregion of the entire map. For each cluster, an individual model is trained to predict the channel gains for unknown locations in its corresponding subregion. 

Assume that $K$ clusters are formed, denoted as $\{ \tilde {\mathcal{S}}_k\subset \tilde {\mathcal{S}} \} _{k=1}^K$. For $\tilde {\mathcal{S}}_k$, the training input is denoted as $\mathcal{X}_k= \{ \mathbf{x}_{i}=\left [ c_{x,\mathrm{bs}},c_{y,\mathrm{bs}},c_{x,i},c_{y,i}\right ]^T |i \in \tilde{\mathcal{S}}_k\}$, and the corresponding output is $\mathcal{E}_k= \{ {e_i|i \in \tilde{\mathcal{S}}_k} \}$, both having the same size of $|\mathcal{X}_k | =| \mathcal{E}_k | =M_k$, where $M_k$ is the number of sample points in cluster $k$. The objective is to find the set of model parameters $\Lambda_k^*$ to predict the channel gains for its corresponding subregion. Denote the predicted training output as $\hat{\mathcal{E}}_k=F_{\Lambda_k}(\mathcal{X}_k)$, where $F_{\Lambda_k}$ describes the mapping rule of the neural network under parameters ${\Lambda_k}$, the parameters are optimized as

\begin{equation}
\label{eq3}
\Lambda_k^*=\arg\min_{\Lambda_k} f_{\mathrm{mse}}(\mathcal{E}_{k}|\hat{\mathcal{E}_{k}})
,\end{equation}
where $f_{\mathrm{mse}}(\cdot )$ denotes the mean squared error (MSE) function.

With all $\Lambda_k^*$ optimized, the composite network for entire CGM construction is described by the union of $F_{\Lambda_k^*}(\cdot)$. 
For an unkown input location point, it is first determined to which cluster (or subregion) it belongs, then the corresponding sub-network is applied to provide channel gain prediction at its location. In the following section, we provide a detailed description of clustering, training, and channel gain prediction, as illustrated in Fig.~\ref{fig1}.

\section{Subregional Learning-based CGM Construction}

\subsection{Clustering-based Data Division}

A data-driven clustering algorithm is employed to divide the sample points in the SCGM into $K$ clusters. Since the sampled data has low dimensionality, we use the K-means clustering algorithm to partition the sample points. Compared to other complex clustering algorithms, such as spectral clustering and deep adaptive clustering, it offers higher computational efficiency and is relatively simple to implement. The clustering is conducted according to the generalized distance between different sample points, calculated using normalized $\mathbf{v}_i$, ensuring equal contribution of each feature to distance calculations. Note that besides the geographical parameters, the channel gain, $e_i$, is also involved in distance calculation. Therefore, when saying that two sample points are close in (generalized) distance, it means not only their geographical locations are close, but also they have similar channel gains. In this case, these points will be classified into the same cluster, and correspondingly, belong to the same geometrical subregion.

The clustering process is detailed as follows. Given a fixed value of $K$, the $K$-means clustering algorithm is adopted, for which we first randomly initialize the clustering centers, expressed as $\mathcal{O}=\left \{ \mathbf{o}_{k}|k=1,...,K, \mathbf{o}_{k}\in \mathbb{R}^{ 5\times 1 } \right \}$. The distance from $\mathbf{v}_{i}\in \tilde {\mathcal{S}}$ to $\mathbf{o}_{k}$ is calculated as $d_{i,k}=\left \| \mathbf{v}_{i}- \mathbf{o}_{k} \right \| $, accordingly, $\mathbf{v}_{i}$ is classified into the corresponding $ \tilde {\mathcal{S}}_k$, where $k=\arg\min_{k}\left \| \mathbf{v}_{i}- \mathbf{o}_{k} \right \|$, and $\left \| \cdot  \right \| $ denotes the Euclidean norm.

When all the sample points in $\tilde{\mathcal{S}}$ have been clustered with the current clustering centers, the sample points in $ \tilde {\mathcal{S}}_k$ are used to update $\mathbf{o}_{k}$, such that 
\vspace{-2mm}

\begin{equation}
\label{eq4}
\mathbf{o}_{k}=\frac{1}{M_{k}}\sum_{i=1}^{M_{k}}\mathbf{v}_{i}
.\end{equation}
After updating $\mathbf{o}_{k}$, all sample points in $\tilde{\mathcal{S}}$ will be reclassified based on the new centers. According to \eqref{eq4}, since the BS is fixed, the division results only correlate with the positions of sample points and their channel gains. This process is iteratively conducted until a new update of $\mathbf{o}_{k}$ does not show significant difference (e.g., the distance between  $\mathbf{o}_{k}$ and its update is less than a pre-defined threshold). Note that there will exist an optimal setting of $K$, which maximizes the predictive efficiency of the model. For instance, a larger $K$ is anticipated in the condition with adequate sample points, and vice versa. 

\vspace{-4mm}
\subsection{Model Training}

Since the channel gains between neighboring locations are similar, there exists spatial correlation among the sample points. However, this spatial correlation diminishes with increasing frequency, primarily due to the reduction in wavelength and the greater susceptibility of signals to blockage, scattering, and reflection in urban environments.\footnote{When the spatial correlation changes, the relevant parameters of the convolutional neural network can be adjusted to capture the local features.} To this end, for each subregion (corresponding to the formed cluster), we construct a one-dimensional convolutional neural network (1D-CNN) model for channel gain prediction. For training, the input $\mathbf{x}_{i}\in \mathcal{X}_k$ is shaped into a two-dimensional feature map defined as $\mathbf{H}_{i}^{(\mathrm{in})}\in \mathbb{R}^{G\times D}$, where $G$ and $D$ denote the number and length of the input feature map, respectively. Then, the obtained input data $\mathbf{H}_{i}^{(\mathrm{in})}$ is sent to the parameter training unit composed of $L$ layers each containing a convolutional and a pooling sub-layer. Following the convolution and pooling operation, $G$ and $D$ are changed accordingly. The prediction results are then obtained through a fully connected (FC) layer. Denoting the predictive channel gains by $\mathbf{\hat{\mathcal{E}}}_k$, the model parameters of the sub-network can be optimized according to \eqref{eq3}. Then, all sub-networks are employed to constitute the modular convolutional neural network (MCNN-1D) model.

\subsection{Channel Gain Prediction}

In the prediction stage, the MCNN-1D model is employed to predict channel gains in the target area. Before prediction, it is essential to determine the subregion to which each test sample belongs. However, for a sample point in the target area, we can only directly obtain its geographical location information. Therefore, the test sample can be categorized into the corresponding subregion based on its location. Specifically, for a test sample defined as $\mathbf{x}_{g}=\left [ c_{x,\mathrm{bs}},c_{y,\mathrm{bs}},c_{x,g},c_{y,g}\right ]^T$, we calculate the geographical distance defined as $d_{g,k}^{\mathrm{geo} }$ between $\mathbf{x}_{g}$ and the clustering center $\mathbf{o}_k$ using the horizontal and vertical coordinates. Then, $\mathbf{x}_g$ belongs to the subregion $k$, where $k$ can be determined by $k=\arg\min_{k}d_{g,k}^{\mathrm{geo}} $. Finally, the predicted channel gain defined as $\hat{e}_g$ is generated by the corresponding sub-network, which can be expressed as $\hat{e}_g =F_{\Lambda_{k}^* }\left ( \mathbf{H}_g^{(\mathrm{in})}  \right )$.

\section{Optimization for the Subregions} \label{Optimization for the Subregions}

In this section, we first find the optimal number of subregions, and then describe the specific process of the uneven subregion sampling and training data reuse strategies.  

\subsection{Optimal Number of Subregions}

Due to the limited number of sample points, the number of subregions exceeding a certain threshold may result in inadequate training data, thereby decreasing prediction accuracy. Therefore, we find the optimal number of subregions based on the performance of the MCNN-1D model. We randomly sample a small subset of sample points from the map as the test set, denoted by $ \mathcal{S}_{\mathrm{te}}=\{\mathbf{v}_{z}|z=1,..., M_{\mathrm{te}}\}$, where $M_{\mathrm{te}}$ represents the number of sample points in the test set. After the training is complete, root mean squared error (RMSE) is used to characterize the prediction accuracy of the $k$-th sub-network, which can be calculated as

\begin{equation}
\label{eq5}
R_k=\sqrt{\frac{1}{M_{\mathrm{te}}^k} {\textstyle \sum_{z=1}^{M_{\mathrm{te}}^k}}({e}_{z}^k-\hat{e}_{z}^k)^{2}}   
,\end{equation}
where $\hat{e}_z^k$ denotes the predicted channel gain and $M_{\mathrm{te}}^k$ denotes the number of sample points in the $k$-th test subset, with $M_{\mathrm{te}}=M_{\mathrm{te}}^1+...+M_{\mathrm{te}}^K$. Correspondingly, the RMSE $R_K$ of the MCNN-1D model with $K$ sub-networks can be calculated as $R_K=\sqrt{\frac{1}{M_{\mathrm{te}}} {\textstyle \sum_{z=1}^{M_{\mathrm{te}}}}({e}_{z}-\hat{e}_{z})^{2}}$. To determine the optimal number of subregions $K^*$, we need to compare the accuracy of the MCNN-1D model for different partitioning numbers. Specifically, $K^*$ is defined as the $K$ corresponding to the lowest $R_K$.

\begin{algorithm}[t]\footnotesize
    \renewcommand{\algorithmicrequire}{\textbf{Input:}}
	\renewcommand{\algorithmicensure}{\textbf{Output:}}
	\caption{Uneven subregion sampling and training data reuse strategies.}
    \label{Algorithm 1}
    \begin{algorithmic}[1] 
        \REQUIRE Clusters $\left \{ \tilde{\mathcal{S}}_1,\dots , \tilde{\mathcal{S}}_{K^*}\right \} $, clustering centers $\left \{ \mathbf{o}_1,...,\mathbf{o}_{K^*} \right \} $, distance threshold $\sigma $. 
	\ENSURE Sampling rate $\lambda _k$, optimized clusters. 
        \FOR {$k=1$ to $K^{*}$}
            \STATE Calculate $R_k$ according to \eqref{eq5}.
        \ENDFOR
        \FOR {$k=1$ to $K^{*}$}
            \STATE Obtain $\lambda _k$ according to \eqref{eq6}.
        \ENDFOR  \COMMENT{Uneven subregion sampling strategy.}
        \FOR {$k=1$ to $K^{*}$}
            \STATE Calculate the average distance $d_{k}$ for $\tilde{\mathcal{S}}_k$.
            \STATE $N_{k,\mathrm{out}}=\left | \tilde{\mathcal{S}}\setminus \tilde{\mathcal{S}}_k \right |$.
            \FOR{$\beta =1$ to $N_{k,\mathrm{out}}$}
                \STATE $d_{\beta ,k}=\left \| \mathbf{v}_{\beta}-\mathbf{o}_{k} \right \|$. \COMMENT{$\mathbf{v_{\beta }}\in \tilde{\mathcal{S}}\setminus \tilde{\mathcal{S}}_k$}
                \IF {$ d_{\beta ,k}-d_{k}< \sigma $} 
                    \STATE $\tilde{\mathcal{S}}_{k}\gets \tilde{\mathcal{S}}_{k}\cup \mathbf{v}_{\beta}$
                \ENDIF
            \ENDFOR
        \ENDFOR  \COMMENT{Training data reuse strategy.}
    \STATE return $\left \{ \lambda _1,\dots , \lambda_{K^*}\right \} $ and $\left \{ \tilde{\mathcal{S}}_1,...,\tilde{\mathcal{S}}_{K^*} \right \} $
    \end{algorithmic}
\end{algorithm}

\vspace{-2mm}
\subsection{Optimization after Training}

In practice, to improve the accuracy of the predicted CGM, additional sampling on the map is a commonly adopted approach. However, subregions with complex propagation environments often require more training data. As shown in Fig. 1, more sample points are selected in $\mathcal{C}_1$ than $\mathcal{C}_2$. When further sampling is conducted, different sampling rates should be applied to the subregions to provide more training data for sub-networks in more complex environments. Additionally, sample points located at the boundaries of each cluster usually contain information from neighboring clusters. These points (denoted as reusable sample points) can be incorporated into different clusters to enrich the training data.

The detailed processes of the two optimization approaches are described in Algorithm \ref{Algorithm 1}. As discussed above, the sampling rates should be determined first. The method proposed in \cite{9587242} calculated the sampling rate based on the size and density of each cluster. In this paper, we further incorporate additional factors, including the accuracy of each sub-network $R_k$ and the variation in channel gains within the clusters $\delta_k$, to enhance the accuracy of the sampling rate calculation. Therefore, referring to the calculation formula in \cite{9587242}, the sampling rate for the $k$-th subregion can be calculated as

\begin{equation}
\label{eq6}
\lambda_k=\frac{\theta_k \delta_k R_k }{ {\textstyle \sum_{q=1}^{K^*}\theta_q \delta_q R_q }} 
,\end{equation}
where $\theta _k=\frac{M_k}{M_{\mathrm{scgm}}} $ denotes the cluster size. And $\delta_k$ represents the variance of channel gains in the $k$-th subregion, calculated as $\delta_k = \frac{ {\textstyle \sum_{i=1}^{M_k}(e_i-\overline{e})^{2}} }{M_k}$, where $\overline{e}$ denotes the average channel gain. Then, we can select $N_k$ sample points in the subregion $k$ based on its sampling rate during the sampling process later, and $N_k$ can be calculated as $N_k=N \times  \lambda _k$, where $N$ denotes the total number of the new sample points within the entire region. For the training data reuse strategy, we determine whether a sample point provides the boundary information based on its generalized distance to the cluster centers. Specifically, we first obtain the average distance $d_k$ from the sample points in the corresponding cluster to the center. Then, we identify the sample points outside cluster $k$, and calculate the distance $d_{\beta ,k}$ from each point $\mathbf{v}_{\beta}$ to the $k$-th clustering center. As shown in Steps 12 and 13 of Algorithm \ref{Algorithm 1}, if the difference between $d_{\beta ,k}$ and $d_k$ is smaller than the distance threshold $\sigma$, the sample point $\mathbf{v}_{\beta}$ will be added to the corresponding cluster.

\section{Simulation Results and Analysis} \label{Experiment Results and Analysis}

\subsection{Simulation Setting}

In this work, we consider a geographic scenario of size $470\times 630$ m$^2$, and the ground truth CGM is obtained by ray tracing, where black areas represent the buildings without setting sample points (as depicted later in Fig.~\ref{fig4}\subref{Channel gain map}). The entire simulation process is implemented on the WinProp platform. The BS equipped with an omnidirectional antenna is located at $[120,540,164]^T$, with the transmit power of $1$ W and carrier frequency of $4,800$ MHz. The sampling interval and height are set to $1$ m and $1.5$ m, respectively. Finally, the entire map contains $24,730$ sample points. We construct the $\tilde{\mathcal{S}}$ via random sampling, where data is collected at random locations.  

Each sub-network includes a convolutional layer and a pooling layer, with convolutional kernels of size ($3\times 1$) and pooling kernels of size ($2\times 1$). The corresponding numbers of kernels are $Z_{conv}=64$ and $Z_{pool}=64$, respectively. After the convolutional and pooling operations, the FC layer is equipped with $64$ neurons. The Adam optimizer with a learning rate $\zeta=1e^{-3}$ is adopted. The whole model is trained for $1,000$ epochs and the batch size is set to different numbers based on the size of $\tilde{\mathcal{S}}$. 

\subsection{Prediction Performance and Analysis}

Table~\ref{tab1} shows the RMSE values of the MCNN-1D model with varying numbers of sub-networks. Since data division results in reduced training sample points in each subregion, the prediction model can obtain the optimal performance without classification when $M_{\mathrm{scgm}}=200$ \cite{10530520}. As $M_{\mathrm{scgm}}$ increases, the data division approach improves the accuracy of map construction, which is consistent with the conclusion found in \cite{7997073}. Specifically, different sub-network models provide unique prediction information for sample points in different propagation environments (such as points $\mathbf{v}_1$ and $\mathbf{v}_2$ in Fig~\ref{fig1}). In addition, the $K^{*}$ also varies with the increase in training sample points.

Note that propagation environments vary across different wireless communication scenarios, and each scenario has its own optimal number of subregions. Intuitively, a large $K$ is expected in scenarios with more complex propagation environments. In practice, the range of possible $K^*$ values can be constrained using the results of Table~\ref{tab1}. Then, the proposed framework enables the rapid determination of the appropriate $K^*$ for a given scenario. In addition, when the environments change, sampling can be performed according to the previously determined partitioning, enabling subsequent prediction tasks to be conducted more efficiently, while saving time and computational resources.

\begin{figure}[t]
\begin{center}
    \includegraphics[width=0.75\columnwidth]{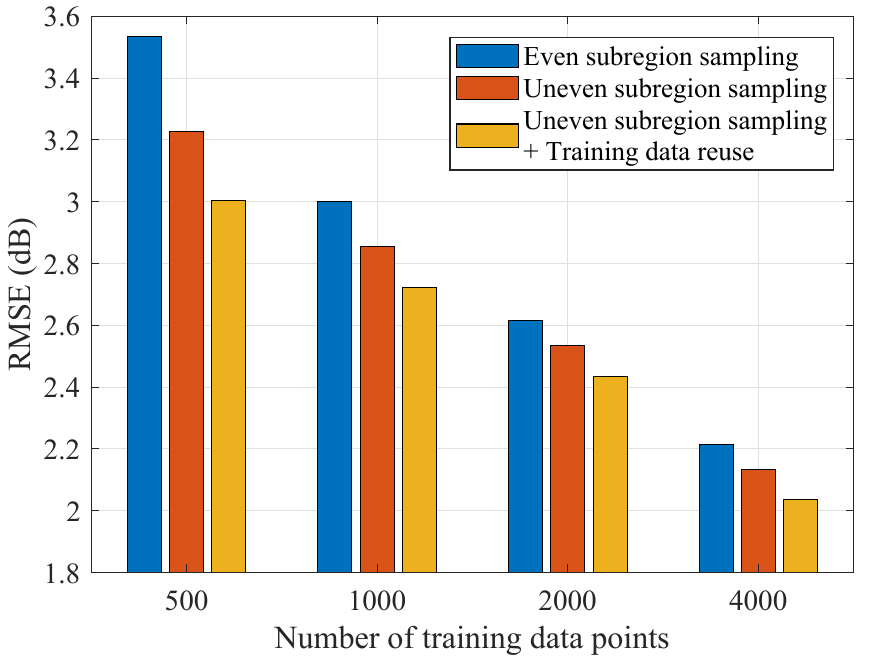}
    \caption{Accuracy of the proposed optimization strategies.}
    \label{fig2}
\end{center}
\vspace{-0.4cm}
\end{figure}

To demonstrate the effectiveness of the two optimization approaches in Algorithm \ref{Algorithm 1}, we first set the number of points for further sampling as $N=0.25 \times M_{\mathrm{scgm}}$. Moreover, the parameter $\sigma$ in the training data reuse strategy is set as $\sigma = 0.5 \times d_k$. Then, we use even subregion sampling as the benchmark, where the number of sample points in each subregion is calculated as $N_k=N\times \frac{1}{K^{*}} $. Fig.~\ref{fig2} illustrates that the uneven subregion sampling strategy improves the prediction accuracy, while training data reuse further enhances accuracy without adding additional training sample points. When the data size increases, there is a significant improvement in prediction accuracy based on the even subregion sampling. However, both two optimization strategies can still enhance the performance of the prediction model to a certain extent. 

\begin{table}[t]\footnotesize
    \begin{center}
    \caption{RMSE of the MCNN-1D model with different numbers of subregions.}
    \label{tab1}
    \setlength{\tabcolsep}{1mm}
    \renewcommand{\arraystretch}{1}
    \begin{tabular}{cccccccccc}
        \toprule
        \multirow{2}*{\shortstack[c]{Data size}} & \multicolumn{9}{c}{Number of subregions} \\
        \cmidrule(lr){2-10}
            & 1 & 2& 3 & 4 &5 & 6 & 7 & 8 & 9 \\
        \midrule
        200 & \pmb{3.912} & 4.074 & 4.108 & 4.363 & 4.262 & 4.437 & 4.235 & 4.398 & 4.818 \\
        400 & 3.765 & 3.693 & \pmb{3.545} & 3.846 & 3.913 & 3.679 & 3.841 & 3.781 & 4.044 \\
        800 & 3.287 & 3.254 & 3.345 & 3.242 & \pmb{3.104} & 3.221 & 3.257 & 3.329 & 3.322 \\
        1600 & 3.159 & 3.049 & 2.982 & 2.801 & 2.722 & 2.770 & \pmb{2.641} & 2.715 & 2.799 \\
        3200 & 2.738 & 2.840 & 2.737 & 2.627 & 2.571 & 2.559 & \pmb{2.468} & 2.567 & 2.501 \\
        \bottomrule
    \end{tabular}
    \end{center}
\vspace{-0.4cm}
\end{table}

\begin{figure}[t]
\begin{center}
    \includegraphics[width=0.75\columnwidth]{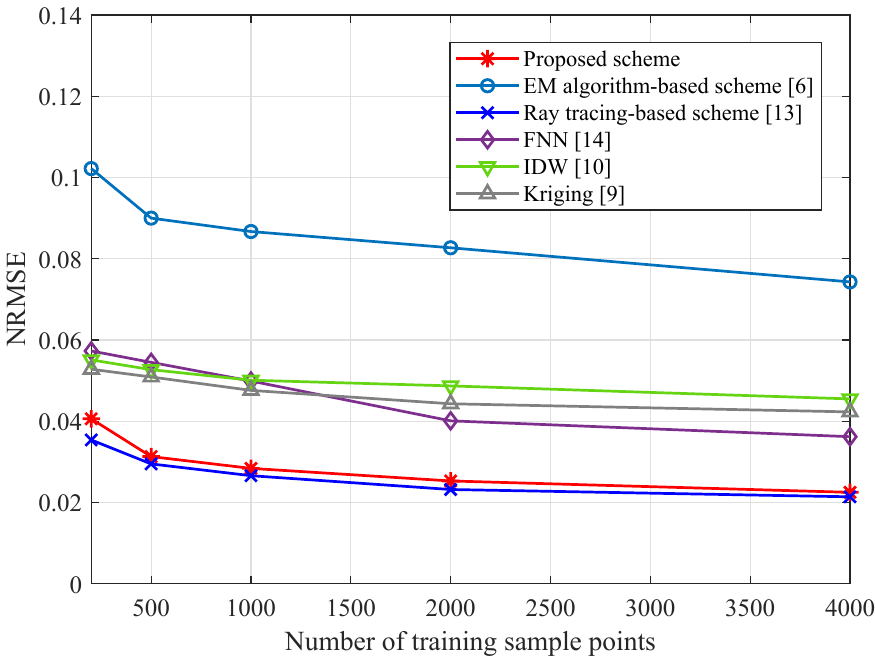}
    \caption{CGM construction with different methods.}
    \label{fig3} 
\end{center}
\vspace{-0.4cm}
\end{figure}

Moreover, we compare the proposed map construction method with both subregion-based and subregion-free approaches. Specifically, the subregion-free approaches include the spatial interpolation techniques (i.e., Kriging \cite{5711699} and IDW \cite{8610487}) as well as the feed-forward neural network approach (FNN \cite{10683332}). For the subregion-based approaches, one benchmark is the EM algorithm-based CGM construction \cite{9771802}. Additionally, while using ray tracing for map partitioning is not feasible for our task, we consider it as the ideal results for map division \cite{8032491}. In this approach, we first divide the sample points into two regions based on LoS and NLoS, and further partition the NLoS region according to the distribution of buildings. Then, the map construction is conducted following the training and prediction methods proposed in this paper. The normalized RMSE (NRMSE) utilized in \cite{9771802} is employed for comparison. As shown in Fig.~\ref{fig3}, due to the complexity of the considered environment, the EM algorithm-based map construction approach may have difficulty in identifying the boundary of each group, which results in prediction errors. In contrast, the proposed CGM construction method outperforms existing approaches and closely approximates the ideal ray tracing-based method.

Next, we compare the computational complexities of three different subregion-based CGM construction approaches. In \cite{9771802}, the computational complexity of the EM algorithm is $O(T \cdot K \cdot M \cdot P)$, where $T$, $K$, $M$ and $P$ represent the numbers of iterations, subregions, sample points and parameters to be estimated, respectively. IDW-based map construction has a computational complexity of $O(M \cdot V)$, where $V$ denotes the total number of sample points in the target area. For ray tracing-based map division, the computational complexity is $O(M \cdot B \cdot I)$, where $B$ and $I$ are the numbers of buildings and rays, respectively. Note that the environmental modeling prior to ray tracing is both time-consuming and challenging to quantify. In the proposed method, the computational complexities for map division and channel gain prediction are $O(T \cdot K \cdot M)$ and $O(V \cdot Z)$, respectively, where $Z$ denotes the total number of operations in the neural network. Additionally, the computational complexity of the training data reuse strategy is $O(K \cdot M)$. Our proposed method has lower computational complexity than the ray tracing-based method. Furthermore, as shown in Fig.~\ref{fig3}, although slightly more complex than the EM algorithm-based method, it achieves a notable improvement in prediction accuracy.
 
\begin{figure*}[t]
\centering
\subfloat[Channel gain map]{
		\label{Channel gain map} \includegraphics[scale=0.2]{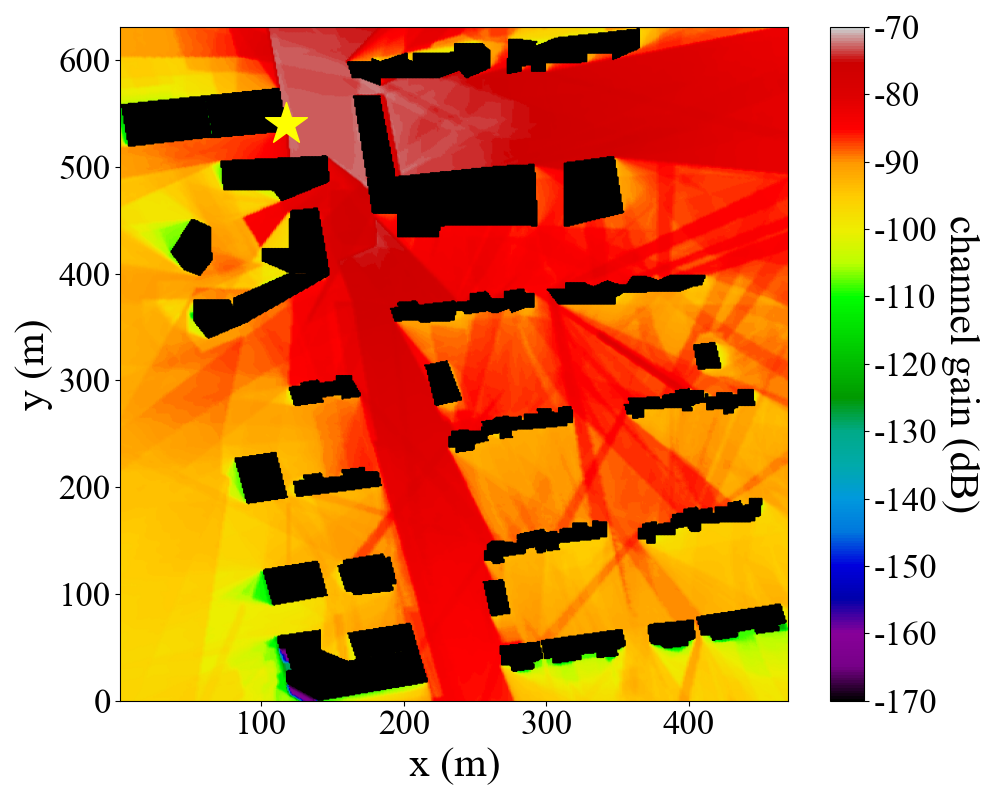}}\hspace{2mm}
\subfloat[Predicted CGM]{
		\includegraphics[scale=0.2]{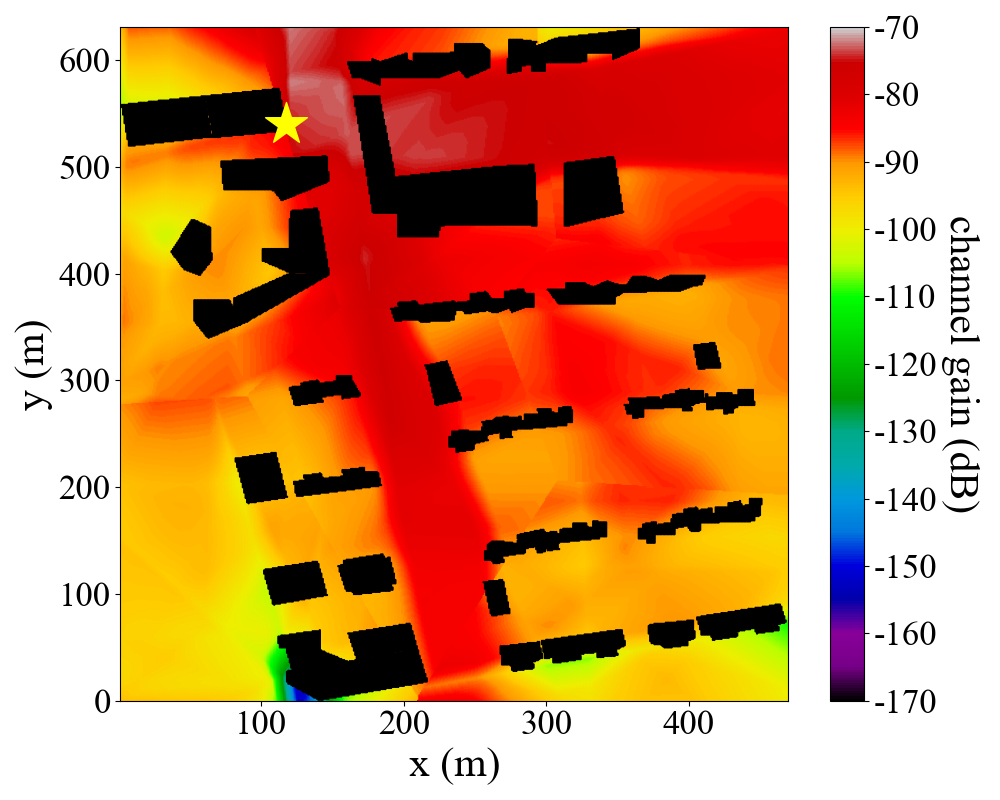}}\hspace{2mm}
\subfloat[Data division results]{
            \label{data division results} \includegraphics[scale=0.285]{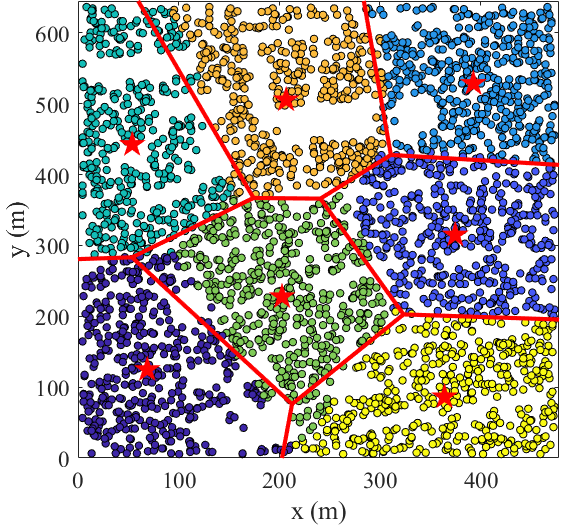}}
\caption{Predicted CGM constructed by the proposed scheme, with $M_{\mathrm{scgm}}=3200$, $N=800$ and $K^{*}=7$.}
\vspace{-3mm}
\label{fig4}
\end{figure*}

To visually demonstrate the effectiveness of the proposed approach, Fig.~\ref{fig4} shows the predicted CGM and the corresponding division results. The areas delineated by the red lines in Fig.~\ref{fig4}\subref{data division results} represent subregions partitioned by geographical distances between sample points and centers (denoted as red stars), while sample points of different colors form clusters divided by generalized distances. It can be observed that a few sample points located at boundaries may be misclassified into other subregions. As indicated by Fig.~\ref{fig2}, the training data reuse strategy allows prediction models to extract richer boundary information, which can reduce prediction errors arising from the factor above. Finally, compared to the ground truth, the proposed map construction method effectively predicts channel gains in target areas. 

The simulation results demonstrate the effectiveness of the CGM construction method proposed in this paper. Note that real-world environments are dynamic, causing continuous changes in channel gains within the map. In practice, we can employ techniques such as transfer learning on the existing model to obtain specific prediction models with minimal additional training. Furthermore, periodic updates to the CGM using real-time measurements or integrating predictive models based on historical data can further enhance the robustness and accuracy of the proposed method in dynamic scenarios.

\vspace{-2mm}
\section{Conclusion} \label{Conclusion}

In this paper, we propose a novel subregional learning-based CGM construction method for realizing environment-aware communications. To distinguish different radio propagation environments in the scenario, we divide the map into various subregions based on a data-driven clustering approach. Subsequently, we construct a deep learning model named MCNN-1D, which trains a sub-network individually for each subregion, and predicts the channel gains for the target areas. To achieve a more precise CGM, we propose uneven subregion sampling and training data reuse strategies to optimize the setup of each subregion. Simulation results corroborate the effectiveness of the proposed scheme.


\end{document}